\documentclass{ws-procs975x65}

\begin{document}



\title{GAMMA-RAY BURST HOST GALAXY GAS AND DUST\footnote{The authors acknowledge funding from the EU RTN `Gamma-Ray
  Bursts: An Enigma and a Tool', support from PPARC, and RS thanks the
  conference organisers for financial assistance.}}

\author{RHAANA STARLING, RALPH WIJERS AND KLAAS WIERSEMA}

\address{University of Amsterdam\\
Kruislaan 403, 
1098 SJ Amsterdam, The Netherlands \\
\email{rlcs1@star.le.ac.uk ; rwijers@science.uva.nl ; kwrsema@science.uva.nl}}


\begin{abstract}
We report on the results of a study to obtain limits on the absorbing columns towards an initial
sample of 10 long Gamma-Ray Bursts observed with {\it BeppoSAX}, using a new
approach to SED fitting to nIR, optical and X-ray afterglow data, in count space and including
the effects of metallicity. When testing MW, LMC and SMC extinction laws we
find that SMC-like extinction provides the best fit in most cases. A MW-like extinction curve is not preferred for any of these
sources, largely since the 2175\AA\ bump, in
principle detectable in all these afterglows, is not present in the
data. We rule out an SMC-like gas-to-dust ratio or lower value for 4 of
the hosts analysed here (assuming SMC metallicity and extinction law) whilst
the remainder of the sample have too large an error to discriminate. We
provide an accurate estimate of the line-of-sight extinction, improving upon the uncertainties for the majority of the extinction measurements made in previous studies of this sample.
\end{abstract}

\bodymatter

\section{Introduction}\label{intro}
The accurate localisation of Gamma-Ray Bursts (GRBs) through their optical and
X-ray afterglows has enabled detailed studies of their environments. Selection
solely by the unobscured gamma-ray flash has allowed the
discovery of a unique sample of galaxies spanning a very wide range
of redshifts from $z$ $\sim$ 0.009 to 6.3\cite{meanz}. Hence, detailed and extensive host galaxy observations provide a
wealth of information on the properties of star-forming galaxies
throughout cosmological history. 

 Afterglow
spectroscopy and/or photometry can be used to provide an estimate of the total
extinction along the line-of-sight to the GRB. Absorption within our own Galaxy along a particular line of sight can be
estimated and removed, but absorption which is intrinsic to the GRB host galaxy as a function of wavelength is unknown, and
is especially difficult to determine given its dependence on metallicity and
the need to distinguish it from that of intervening systems.
In general, low amounts of optical extinction are found towards GRBs,
unexpected if GRBs are located in dusty star-forming regions, whilst the X-ray
spectra reveal a different picture. At
X-ray wavelengths we often measure high values for the absorbing columns, where the absorption is caused by metals in both gas and solid phase\cite{xrayabs}. The apparent discrepancy between optical and X-ray extinction resulting in
high gas-to-dust ratios in GRB host galaxies (often far higher than for the
MW, LMC or SMC, e.g. GRB\,020124\cite{020124}) is not satisfactorily explained, though the suggestion that dust
destruction can occur via the high energy radiation of the GRB\cite{waxdraine} could possibly account for the discrepancy. 

Traditionally the optical and X-ray spectra have been treated seperately in
extinction studies. Since the underlying spectrum is likely a synchrotron
spectrum (power law or broken power law) extending through both wavelength regimes, it is most accurate
to perform simultaneous fits.
We perform simultaneous broadband fits of the spectral energy distributions (SEDs) in count space, so we need not first assume a model for the X-ray spectrum. Inclusion of nIR data
and $R$ band optical data together with the 2--10 keV X-ray data, regions over
which
extinction has the least effect, allows the underlying power law slope to be most
accurately determined. This sample of 10 long GRBs observed with the {\it BeppoSAX} Narrow Field
Instruments is chosen for the good
availability (3 bands or more) of optical/nIR photometry.

\section{Results and Discussion}\label{resdiss}
\begin{figure}
\psfig{file=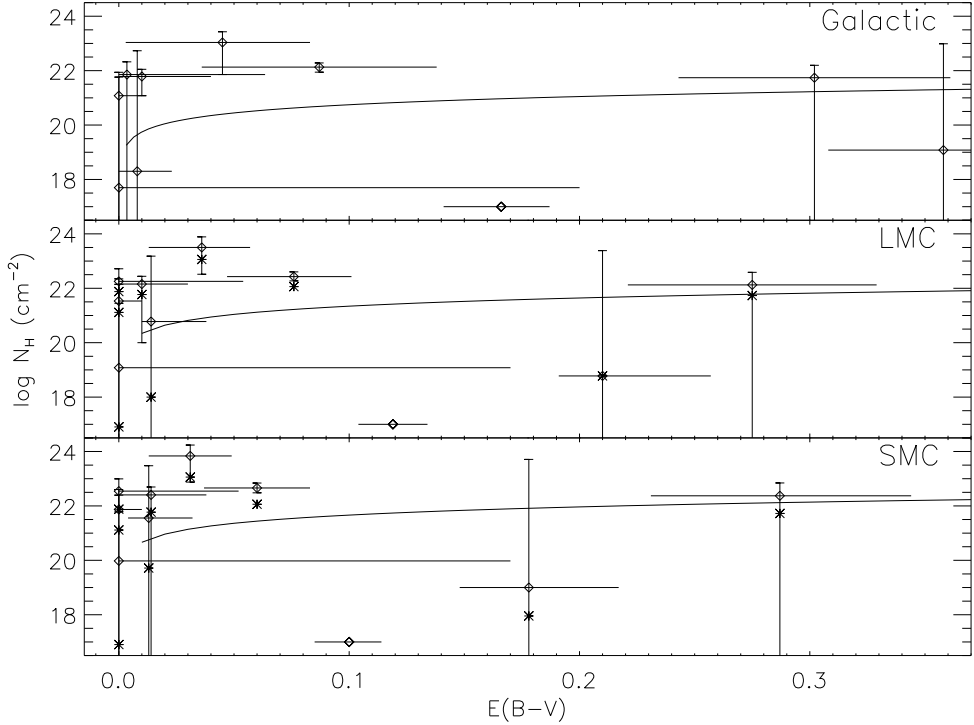,width=3.8in,angle=90}
\caption{Intrinsic absorption in optical/nIR
  ($E(B-V)$) and X-rays (log $N_{\rm H}$) measured for the GRB sample with 90
  \% error bars. We
  compare these with three different optical
  extinction laws overlaid with solid curves: Galactic (top panel), LMC (middle panel) and SMC
  (lower panel). Appropriate
  metallicities are adopted for LMC (1/3 Z$_{\odot}$) and SMC (1/8
  Z$_{\odot}$) calculations (diamonds), and stars mark the
  centroids of the Solar metallicity fits. For
  GRB\,000926 the data were too sparse to fit for $N_{\rm H}$, so we plot the
  $E(B-V)$ range at log $N_{\rm H}$ = 17.0 for clarity.}
\label{fig3}
\end{figure}

Detailed results of fits to the SEDs for all GRBs in the sample, and
further references, can be
found in Starling et al. (2007). Figure~\ref{fig3} shows a comparison of the absorption measurements with Galactic, LMC and SMC
gas-to-dust ratios. This plot has
been constructed in a number of previous works\cite{GalWij01} \cite{stratta}
\cite{kann} \cite{schady} and here we show
the observed distribution of $E(B-V)$ and $N_{\rm H}$ for the first time
derived simultaneously from a fit to X-ray, optical and nIR data. We find a
large excess in absorption above the Galactic values in two sources: GRBs 000926 ($E(B-V)$ only) and 010222, whilst no significant intrinsic absorption is necessary in GRBs 970228 and 990510. The cooling break can be located in three of the afterglows: GRBs 990123, 990510 and 010222 and to all other SEDs a single power law is an adequate fit.

We find a wide spread in
central values for the gas-to-dust ratios, and for 4 GRBs the gas-to-dust
ratios are formally
inconsistent with (several orders of magnitude higher than) MW, LMC and SMC values at the 90 \% confidence limit
assuming the SMC metallicity. This must mean that either gas-to-dust ratios in
galaxies can span a far larger range than thought from the study of local
galaxies, or the ratios are disproportionate in GRB hosts because the dust is
destroyed by some mechanisms (likely the GRB jet), or that the lines of sight
we probe through GRBs tend to be very gas-rich or dust-poor compared with
random lines of sight through galaxies. A dust grain size
distribution which is markedly different than considered here may also affect
these ratios.

We have compared the results of this method to those of other methods of determining
$E(B-V)$. In particular we find that with respect to continuum fitting methods such as this,
optical extinction is overestimated with the depletion pattern method\cite{savfallfiore}, and we have
quantified this for a small number of cases\cite{starling}.
We note, however, that since this is a line-of-sight method,
the measured columns may not be representative of the host galaxy as a whole,
therefore comparison with the integrated host galaxy methods is important.

Swift, robotic telescopes and Rapid Response Mode on large telescopes such as the William Herschel Telescope and the Very Large Telescopes now allow early, high quality data to be obtained, which will help immensely in discriminating between the different extinction laws at work in the host galaxies.

\vfill

\end{document}